\documentclass[aps,pra,superscriptaddress,floatfix,twocolumn,a4paper,showpacs]{revtex4}

\usepackage{graphicx}
\usepackage{amsmath}
\usepackage{amssymb}
\usepackage{epsf,latexsym}
\usepackage{times}

\newcommand{\ket}[1]{\ensuremath{\left| #1 \right\rangle}}
\newcommand{\bra}[1]{\ensuremath{\left\langle #1 \right|}}

\newcommand{\be}{\begin{equation}}
\newcommand{\ee}{\end{equation}}
\newcommand{\ba}{\begin{eqnarray}}
\newcommand{\ea}{\end{eqnarray}}

\newcommand{\Fig}[1]{Fig.~\ref{#1}}

\newcommand{\BUE}{Centre for Theoretical Physics,
               The British University in Egypt,
               El Sherouk City, Postal No. 11837, P.O. Box 43, Egypt.}
\newcommand{\HP}{Hewlett-Packard Laboratories, Filton Road,
              Stoke Gifford, Bristol BS34 8QZ, United Kingdom}
\newcommand{\LBRO}{Department of Physics, Loughborough University,
                Loughborough, Leics LE11 3TU, United Kingdom}

\begin{document}


\title{The quantum-classical crossover of a field mode}

\author{M. J. Everitt}
\email{m.j.everitt@physics.org}
\affiliation{\BUE}
\affiliation{\LBRO}
\author{W, J. Munro}
\affiliation{\HP}
\author{T. P. Spiller}
\affiliation{\HP}

\date{\today }

\begin{abstract}
We explore the quantum-classical crossover in the behaviour of a
quantum field mode. The quantum behaviour of a two-state system---a
qubit---coupled to the field is used as a probe. Collapse and
revival of the qubit inversion form the signature for quantum
behaviour of the field and continuous Rabi oscillations form the
signature for classical behaviour of the field. We demonstrate both
limits in a single model for the full coupled system, for field states
with the same average field strength, and so for qubits with the
same Rabi frequency.
\end{abstract}

\pacs{03.65.-w,03.65.Ta,03.65.Yz,03.67.-a,42.50.-p}


\maketitle

There has long been interest in how fundamental quantum things
give rise to the classical behaviour much more familiar to us
in our everyday lives. Since the time of ``Ehrenfest's
theorem''~\cite{ehr27} it has been known that expectation values
follow their corresponding
classical mechanical equations. However, this isn't the whole story,
and so most of us seek rather more
than just this, since spreading quantum waves hardly look like
classical lumps, or particles.

Quantum measurement~\cite{whezur} has played an extremely important
role in this area. Clearly it isn't possible to observe any
form---quantum, or tending to classical---of behaviour of a quantum
system without measurement. It therefore makes sense to consider the role
of measurement apparatus, or the degrees of freedom that are used
to observe the behaviour of the system of interest. An early example of this
is the work of Neville Mott on $\alpha$-ray tracks~\cite{mott29}.

The role of decoherence in general is also very important in this
area~\cite{zeh,zurek}. Loosely speaking, since measurement
interactions---discrete and projective, or continuous---can be
thought of as localising quantum states (so spreading waves begin
to look more like particles following trajectories), other forms
of decoherence that provide similar localisation effects contribute
to quantum systems behaving more classically. Indeed, within this
picture, in which quantum measurement is regarded as just one
example of decoherence applied to a system, it may be that other
forms of decoherence, such as dissipative coupling to an
environment, are dominant. So these sorts of decoherence alone
may effect the classical limit of a quantum system's behaviour,
whether or not anyone (with or without a PhD~\cite{bell})
is looking.

One explicit example of this is the emergence of chaos in
dissipative quantum systems~\cite{spiral,brun,habib98,everitt1,dyrting96,Bhattacharya00,Habib06,Kapulkin08}.
Non-linear dissipative and driven classical systems can exhibit
chaos, demonstrated, for example, through a periodic phase space
plot showing a strange attractor, as opposed to isolated points
corresponding to regular motion. Various examples have been
given~\cite{spiral,brun,habib98,everitt1,dyrting96,Bhattacharya00,Habib06,Kapulkin08} of the emergence of such
chaos---a characteristic of classical dynamical motion---in
the corresponding dissipative and driven quantum systems. These
examples generally involve scaling the parameters governing the
quantum dynamics so that the phase space motion becomes large on
the scale of $\hbar$ and the quantum state starts to look
like a localised lump following a classical trajectory.
Similar behaviour can emerge in Hamiltonian
chaotic systems, suitably measured~\cite{everitt2}.

In the work presented here, we focus on the example of an
oscillator, or field mode. We'll show how both characteristic
classical and quantum behaviour can emerge from the single framework
of a dissipative quantum model, as appropriate parameters are
varied. In order to achieve this, we couple a two-level quantum
system/atom, or qubit, to the
field mode. In effect, the qubit is used to monitor the regime of
behaviour of the field. This simple model is applicable to a broad
range of physical systems---atomic or fabricated qubits coupled to
an optical cavity; atomic or charge qubits coupled to a microwave
cavity; charge or magnetic qubits coupled to a nanomechanical
resonator.

\begin{figure}[b]
\begin{center}
\resizebox*{0.48\textwidth}{!}{\includegraphics{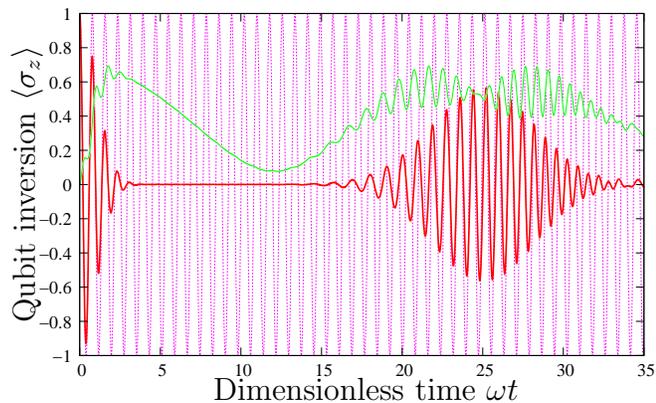}}
\end{center}
\caption{(Color online) Qubit inversion, $\langle \sigma_z \rangle$,
as a function of dimensionless time $\omega t$ for the resonant
cases of Rabi oscillations (dotted; light grey/magenta) in a classical field and collapse and
revival (dark grey/red heavy line) in a quantum field. For the classical field case
$\nu/\omega = 8$ is used, corresponding to the same dominant Rabi frequency
(see \cite{gerkni}) as in
the quantum field case, which uses a coherent state
(\ref{coh}) with $\alpha = \sqrt{15}$ and $\lambda/\omega = 1$. Also shown (in green/light grey solid line)
is the qubit entropy (in Nats), which indicates the degree of entanglement with the field.
}
\label{rabirev}
\end{figure}

The classical limit of the field behaviour is demonstrated
through continuous Rabi oscillations of the qubit~\cite{rabi37,gerkni}.
A qubit, driven by an appropriate oscillatory {\it classical} field,
oscillates between its two energy eigenstates. If these are
taken to be the eigenstates of the $\sigma_z$ Pauli matrix,
separated by energy $\hbar \omega_0$, an appropriate Hamiltonian
for the qubit coupled to an external classical field of frequency
$\omega$ is
\be
\label{Hrabi}
H_{cl} = \frac{\hbar \omega_0}{2} \sigma_z + \hbar \nu \cos \omega t \:\sigma_x \; .
\ee
The detuning is defined as $\Delta=\omega_0-\omega$ and the Rabi
oscillation frequency as $\Omega_R=\sqrt{\Delta^2 + \nu^2}$,
so for the case of the field on resonance with the qubit
(zero detuning, $\Delta=0$) the Rabi frequency is simply $\Omega_R = \nu$. It
is set by the amplitude of the field, not its frequency. In this
resonant case, a qubit initially in state $\ket{\uparrow_z}$ oscillates
fully to $\ket{\downarrow_z}$ at frequency $\Omega_R$. In the language of
atomic and optical physics, the atomic inversion---in qubit language
$\langle \sigma_z \rangle$---satisfies $\langle \sigma_z \rangle =
\cos \Omega_R t$. In the absence of any decoherence acting on the qubit,
these Rabi oscillations persist and are a well known characteristic of
a qubit resonantly coupled to an external classical field. An example
is shown in figure \ref{rabirev}.

The quantum limit of the field behaviour is demonstrated through
collapse~\cite{cumm65,sten73,meys75} and revival
\cite{ebe80,naro81,kni82,gerkni} of the qubit Rabi oscillations.
If, instead of treating the field as a classical source in the Hamiltonian
of the qubit, the field is treated as a quantum oscillator mode
Jaynes-Cummings Hamiltonian~\cite{jaycumm63,gerkni} provides the appropriate description
\be
\label{Hjc}
H_{q} = \frac{\hbar \omega_0}{2} \sigma_z +  \hbar \omega \left(a^{\dagger}a+\frac{1}{2}\right)
+ \hbar \lambda \left(\sigma_{+}a+\sigma_{-}a^{\dagger}\right) \;.
\ee
Here $a^{\dagger}(a)$ is the creation (annihilation) operator for the field, with $[a,a^{\dagger}]=1$,
$\sigma_{\pm}=\frac{1}{2}\left(\sigma_x \pm i \sigma_y\right)$ are the qubit operators that effect
transitions between the energy ($\sigma_z$) eigenstates, and $\hbar \lambda$ is the coupling energy
between the qubit and the field.  In order to derive the Rabi oscillations in a classical
field analytically, it is usual to chuck away the rapidly rotating terms. This (very good) rotating wave
approximation is also made in the derivation of equation (\ref{Hjc}) from the underlying dipole interaction
between the qubit and field, so both quantum and classical models are on the same footing.
In the quantum case, a qubit initially in state $\ket{\uparrow_z}$ is coupled to a
coherent state of the field
\be
\ket{\alpha}=e^{-|\alpha|^{2}/2} \sum_{n=0}^{\infty} \frac{\alpha^n}{\sqrt{n!}}\ket{n} \; ,
\label{coh}
\ee
where the basis $\{\ket{n}\}$ comprises the Fock, or number, states of
the field.  We note that $\ket{\alpha}$ is an eigenstate of the
annihilation operator labelled by its eigenvalue $\alpha$.  In this
scenario, although the qubit initially exhibits Rabi oscillations
analogous to those in the classical case, these apparently ``decay'',
and then subsequently revive~\cite{ebe80,naro81,kni82,gerkni}.  An
example is shown in figure~\ref{rabirev}. Such collapse and revival of
Rabi oscillations of a qubit is widely recognised as a characteristic
of a qubit coupled to a quantum field mode. It is understood both
theoretically~\cite{geaban90,geaban91,knibuz95} and
experimentally~\cite{auf03,meu05} that the apparent decay of qubit
coherence is due to entanglement with the field mode, generated by the
coherent evolution of the coupled quantum systems. 
This is illustrated in
figure~\ref{rabirev} through a plot of the qubit entropy, $S(t)= -
\rm{Tr} (\rho(t) \rm{ln} \rho(t))$ where $\rho(t)$ is the reduced
density matrix of the qubit resulting from a trace over the field (or
vice versa, given the initial system state is pure). Clearly there is
a sharp rise in entropy sympathetic with the initial collapse. The
qubit then disentangles from the field at the ``attractor
time''~\cite{geaban90,geaban91}, half way to the revival. The revival
arises through oscillatory re-entanglement of the qubit and field, as
seen through the subsequent entropy oscillations that coincide with
the revival. Obviously there is no entanglement between the qubit and
the field in the classical limit, because the field is classical, so
the qubit entropy is zero for all times.

It is well known that there are three timescales in the collapse and revival situation
\cite{gerkni}. For fields with
large $\langle n \rangle = \bar{n}$, the Rabi time (or period) is given by
$t_R=2 \pi \Omega_{R}^{-1}=\pi/\left(\lambda \sqrt{\bar{n}}\right)$, the collapse time
that sets the Gaussian decay envelope of the oscillations by $t_c= \sqrt{2}/\lambda$
and the (first) revival time that
determines when the oscillations reappear, such as in the example of figure \ref{rabirev},
by $t_r=2 \pi \sqrt{\bar{n}}/\lambda$.
For the coherent state (\ref{coh}) the average photon, or excitation, number is
$\bar{n} = |\alpha|^2$. Note that the different dependencies of the times on $\sqrt{\bar{n}}$
(which corresponds to the---e.g. electric---field strength of the coherent
field state) allow a sort of
``classical limit'' to be taken. As $\bar{n}$ is increased, there are more Rabi periods packed
in before the collapse---so this {\it appears} more like persistent Rabi oscillations---and the
revival is pushed out further in time. However, the collapse (and revival) still occur eventually,
and in any case the reason there are more Rabi oscillations before the collapse is due to the
inverse scaling of $t_R$ with $\sqrt{\bar{n}}$, so the actual Rabi period is shortened as
$\bar{n}$ is increased. The classical limit we consider in this work is quite different.
We shall consider a fixed $\bar{n}$, so the Rabi period of the qubit does not change in
our various examples. We'll show how the transition from quantum (collapse and revival) 
to classical (continuous Rabi oscillations) can
be effected by introducing decoherence to the quantum field. Our work complements the
dissipative, small-$n$, short-time study of 
Kim et al. \cite{kim99}, who show that such fields are sufficiently classical to
provide Ramsey pulses to Rydberg atoms.

Next we introduce and discuss the decoherence applied to the quantum field model,
used to push it into
a regime of classical behaviour. We use the very simple model---it's all we need to make the
demonstration---of a Lindblad~\cite{lin76} or Bloch type master equation. Here, the evolution of the
density operator of the quantum system of interest, $\rho(t)$, depends only on $\rho(t)$
and not its history, so there are no memory effects (i.e. the evolution is Markovian),
\be
\dot{\rho}=-\frac{i}{\hbar}\left[H_q,\rho\right] + \sum_m \left(L_m \rho L_{m}^{\dagger}
- \frac{1}{2}L_{m}^{\dagger} L_m \rho -\frac{1}{2} \rho L_{m}^{\dagger} L_m \right) .
\label{lindblad}
\ee
The evolution generated by equation (\ref{lindblad}) is irreversible and non-unitary. The
first term is the conventional Schr\"{o}dinger evolution, but the terms due to the
operators $\{L_m\}$ introduce the irreversibility. These can be thought of as the
(approximate) by-product of coupling the quantum system of interest to an environment---a
bath of other quantum degrees of freedom. This environment introduces decoherence in
the evolution of the (reduced) density operator of the quantum system of interest.

In our case, we introduce damping to the quantum field, but no direct decoherence to
the qubit. So we use just a single operator, $L=\sqrt{\gamma} a$, where $\gamma$ is
the decay constant. Implicitly we work at zero temperature, although
the work could easily be extended to finite temperature (introducing thermal noise
on top of quantum noise in the environment) by use of a second operator proportional
to $a^{\dagger}$ and suitably chosen (temperature-dependent) coefficients of
both operators. This damping form of decoherence is quite generic.
Examples of model environments that give such damping
are a bath of two-level atoms (qubits), or a bath of harmonic oscillator
modes. For a field mode in a cavity, these environments respectively represent
a lossy atomic medium in the cavity, or external modes to which the field can leak. For explicit
derivations of equation (\ref{lindblad}), giving damping for a field mode
coupled to such baths, see reference~\cite{ssl}. 

In order to explore the full spectrum of quantum through to classical behaviour of
the field, at a fixed Rabi frequency, we need to use a field strength which is
significantly larger than the simple examples of figure \ref{rabirev}. The field
strength has to be large compared to the quantum uncertainty in the field, so it
can be pushed into a good classical limit, which requires $\bar{n} \gg 1$. However
the field strength cannot be so large that the collapse and revival computations in
the quantum limit become intractable. We have therefore used $\bar{n}=50$ for
the examples presented here. A further constraint is the use of weak coupling. In
the usual discussions of collapse and revival~\cite{gerkni}, the coupling $\lambda$ in
equation (\ref{Hjc}) simply scales the time axis, as all three times $t_R$,
$t_c$ and $t_r$ scale as $\lambda^{-1}$. However, when decoherence is applied to
the field mode, in order to get clean signatures from our qubit probe, it is desirable
to couple it weakly to the field. Also, thinking perturbatively, the back-reaction of
the qubit on the field has to be weak to enable the field to approach the limit of
a fixed classical source in our driven dynamical model.
We therefore use weak coupling of $\lambda/\omega = 0.0005$.

The large $\bar{n}$ and weak coupling regime in which we work equates to some
pretty heavyweight computing. Rather than solving the master equation (\ref{lindblad})
directly, which would be a substantial matrix evolution for a very long time,
we therefore choose to solve an unravelling state evolution equation for the
system, equivalent to (\ref{lindblad}). This only requires solution of state vector,
rather than density matrix, evolution. The unravelling we use is quantum state
diffusion (QSD)~\cite{gisper92,per}. In an unravelling description, a quantum
state $\ket{\psi(t)}$ is used, such that the density operator is recovered
from $\rho(t) = M(\ket{\psi(t)}\bra{\psi(t)})$ where $M$ denotes the mean over
an ensemble. For QSD, the state evolves according to
\ba
\ket{d\psi} = -\frac{i}{\hbar}H_s \ket{\psi} dt
+ \sum_m \left(L_m - \langle L_m \rangle_{\psi} \right) \ket{\psi} d\xi_m
+ \nonumber \\
\sum_m \left(\langle L_{m}^{\dagger} \rangle_{\psi} L_m - \frac{1}{2} L_{m}^{\dagger} L_m
- \frac{1}{2} \langle L_{m}^{\dagger} \rangle_{\psi} \langle L_{m} \rangle_{\psi} \right)
\ket{\psi} dt
\label{qsd}
\ea
where the operators are defined as in (\ref{lindblad}) and the $d\xi_m$ are
independent complex differential random variables satisfying
$M(d\xi_m)=M(d\xi_m d\xi_n)=0$ and
$M(d\xi_m^* d\xi_n)= \delta_{mn}$.
The evolution of (\ref{qsd}) is equivalent to that of (\ref{lindblad}) when the
ensemble mean over the noise is taken. However, it is also the case that in
certain situations where the effects of the noise are small, such as the quantum limit
where the decoherence is very weak, or a classical limit where the motion
is large compared to the noise, a single run of a state unravelling
essentially provides a reasonable approximation to the full ensemble average,
at least for non-chaotic systems.
We utilise this for our work here, as these limits are what interest us and, 
for the parameters we employ, full ensemble averaging out to and 
past the revival time is computationally infeasible.

In order that the quantum field always follows an evolution that gives rise
to our chosen Rabi frequency for the probe qubit, as we add damping to the
field to make it more classical, we also need to add an
external  drive term to maintain
it, so it continues to oscillate with the chosen value of $\bar{n}$. For damping
with a decay constant $\gamma$, this requires a drive term applied to the
field of
$H_d = -\hbar \gamma \alpha \cos \omega t \: \left(a + a^{\dagger}\right)$.
As the field is always started in the coherent state $\ket{\alpha}$, there are
no transients as it settles down to its damped, driven trajectory. In the limit
that there is no decoherence acting, there is also no drive, and we shall recover
the quantum limit of collapse and revival. 

Before presenting our results, we note two further points. \\
(i) Introducing damping through the master equation does not
produce the expected shift in oscillator frequency, which for our
definition of $\gamma$ is
$\omega \rightarrow \sqrt{\omega^{2} -\frac{\gamma^2}{4}}$.
Therefore, in order to ensure the correct behaviour of the field mode in the presence of 
damping we also need to add a further term to the Hamiltonian~(\ref{Hjc}):
$\frac{\hbar \gamma i}{4}\left({a^\dag} ^2 -a^2\right)$. We note that in the 
literature this term is usually expressed in terms of position $q$ and 
momentum $p$ operators in the form $\frac{\gamma}{4}(qp+pq)$, (see for example~\cite{brun,per}). \\
(ii) Detailed studies have been made (see for example~\cite{dae92,geaban93}) of the Jaynes-Cummings system with dissipation applied to the field, resulting in photon loss. For our investigation of the classical limit, we require no net photon loss (to maintain the chosen Rabi frequency). Therefore it is crucial that we also include a suitable drive term, to compensate and prevent damping of the field. This driving is, of course, what is done in experiments in order to apply a constant amplitude classical field to a quantum system using a lossy resonator.

\begin{figure*}
\begin{center}
\resizebox*{1.0\textwidth}{!}{\includegraphics{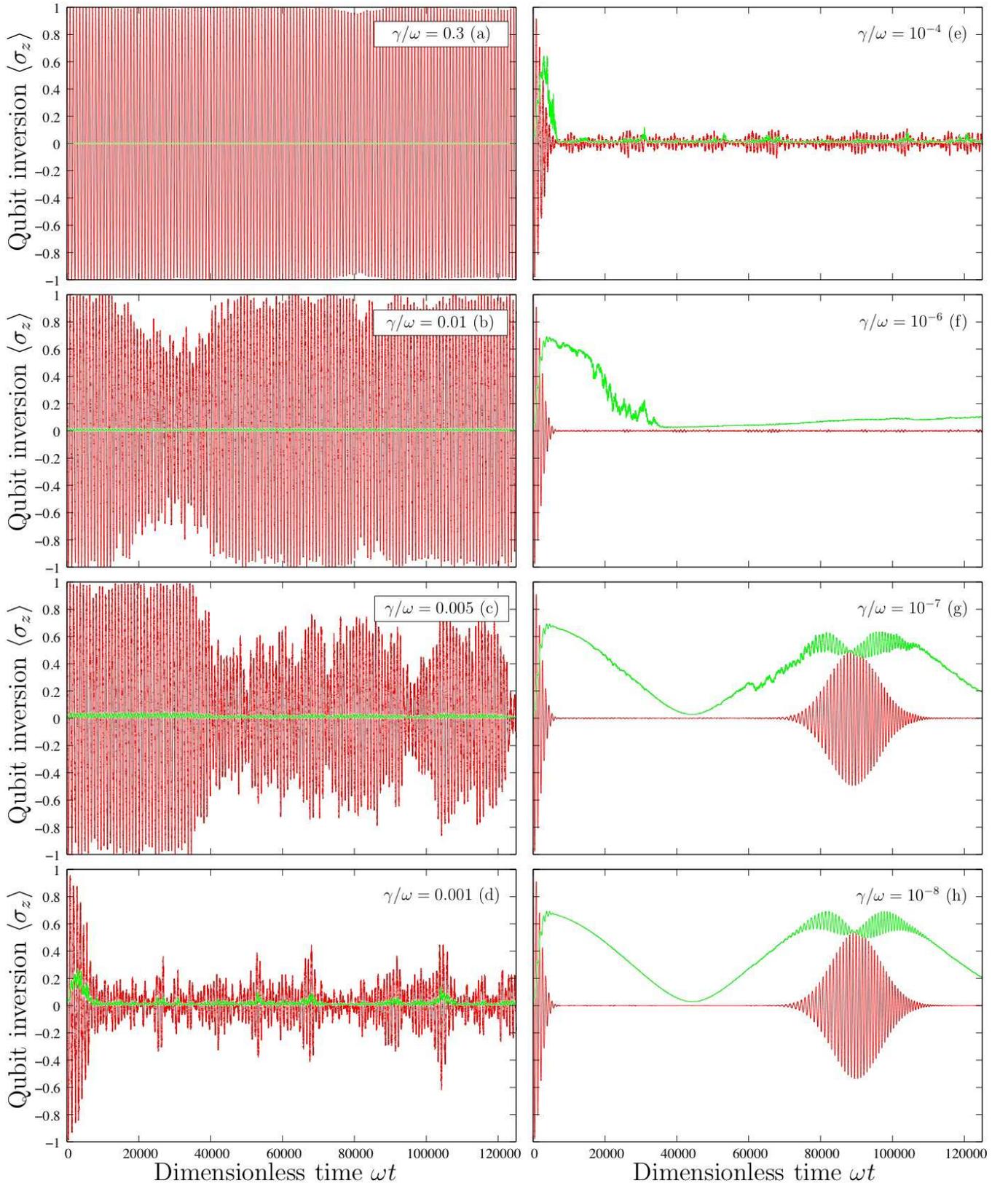}}
\end{center}
\caption{(Color online) Results (dark grey/red) showing 
qubit inversion, $\langle \sigma_z \rangle$,
for different values of dissipation (and therefore drive) applied to the field.
These illustrate
collapse and revival, suppression of collapse and an approach to complete
Rabi oscillations as dissipation and drive are increased. For each
individual QSD run shown, the qubit entropy in Nats (light grey/green) is superimposed.} \label{fig2}
\end{figure*}
So now to our results. In figure~\ref{fig2} we show the qubit inversion
$\langle \sigma_z \rangle$ for various values of damping (and thus also drive)
applied to the quantum field, each computed using a single run of 
QSD. For the
field part of the system to behave classically its coherence time should be short
compared to the Rabi time, so $\gamma \gg \lambda \sqrt{\bar{n}}/\pi$ ($\sim 10^{-3}$ 
for our parameters), as
in the top left plot. For the field to approach the quantum limit, its
coherence time should exceed the revival time, so
$\gamma \ll \lambda/\left(2 \pi \sqrt{\bar{n}}\right)$ ($\sim 10^{-5}$ for our
parameters), as in
the bottom right plot. The pure undamped collapse and revival behaviour is
accurately reproduced in this limit.
In both the quantum and classical limits, just a single run of QSD gives a
very reasonable representation of the qubit
inversion, because the effect of the quantum noise is small.
The intermediate plots are in the quantum-classical crossover region.
Here the quantum noise is significant---a full ensemble average 
(which is beyond our numerical capability) would be needed to
give the actual qubit inversion. Nevertheless, even in these example individual QSD runs
there is evidence of the qubit Rabi oscillations extending beyond the basic
collapse of the quantum limit, as the decoherence of the field is increased.

In our approach, there are two ways in which the qubit state could become mixed.
Firstly, it could entangle with the field~\cite{geaban90,geaban91,knibuz95,auf03,meu05},
as happens in the pure quantum limit to generate the collapse.
Such entanglement can be inferred from the qubit entropy in a single run of QSD (for which
the full qubit-field state is pure). Secondly, the qubit could remain pure in an individual QSD
run, but, when averaged over an ensemble, show mixture. For the classical limit of the top left
plot of figure~\ref{fig2}, we have calculated that both of these effects are small for times
in excess of the collapse time. The qubit-field entanglement remains very close to zero for all
times in a single QSD run, as shown in the entropy plot presented. Independent QSD runs have been
made and these show that 
the qubit mixture is still very small at the collapse time.
Therefore the persistence of good Rabi oscillations well beyond the collapse time and all the
way out to the revival time,
as illustrated in the top left plot of figure~\ref{fig2}, provides a clear signature of
the classical limit of the field. In this limit, the quantum field state is a localized lump 
in phase space (like a coherent state), following the expected classical trajectory and
suffering negligible back-reaction from the qubit. However, the field coherence time is so short
as to prevent entanglement with the qubit developing, unlike in the 
quantum limit~\cite{geaban90,geaban91,knibuz95,auf03,meu05}.
The resultant qubit Rabi oscillations are thus like those due to a classical field,
and not like those \cite{gerkni} that arise from entanglement with a single Fock state 
(which is a delocalized ring in phase space).

Further insight into the classical limit can be obtained from the
phase space behaviour of the field. It is well known that in the pure
quantum limit~\cite{geaban90,geaban91,knibuz95,auf03,meu05} the qubit-field
entanglement correlates distinct and localised (coherent-like) states
of the field with different qubit amplitudes. Thus, when there is
no entanglement at the ``attractor time'', the interaction of the atom
with the field mode generates a
macroscopically distinct superposition of states in the
oscillator---a Schr\"{o}dinger cat state. As one would expect,
and in order to render the field behaviour classical,
the introduction of decoherence suppresses this
phenomenon. We illustrate this in~\cite{mov} by providing two
animations of the dynamics of the Wigner function and atomic inversion
for the parameters of \Fig{rabirev}, one undamped and one with
dissipation of $\gamma/\omega=0.01$.

\begin{figure}
\begin{center}
\resizebox*{0.48\textwidth}{!}{\includegraphics{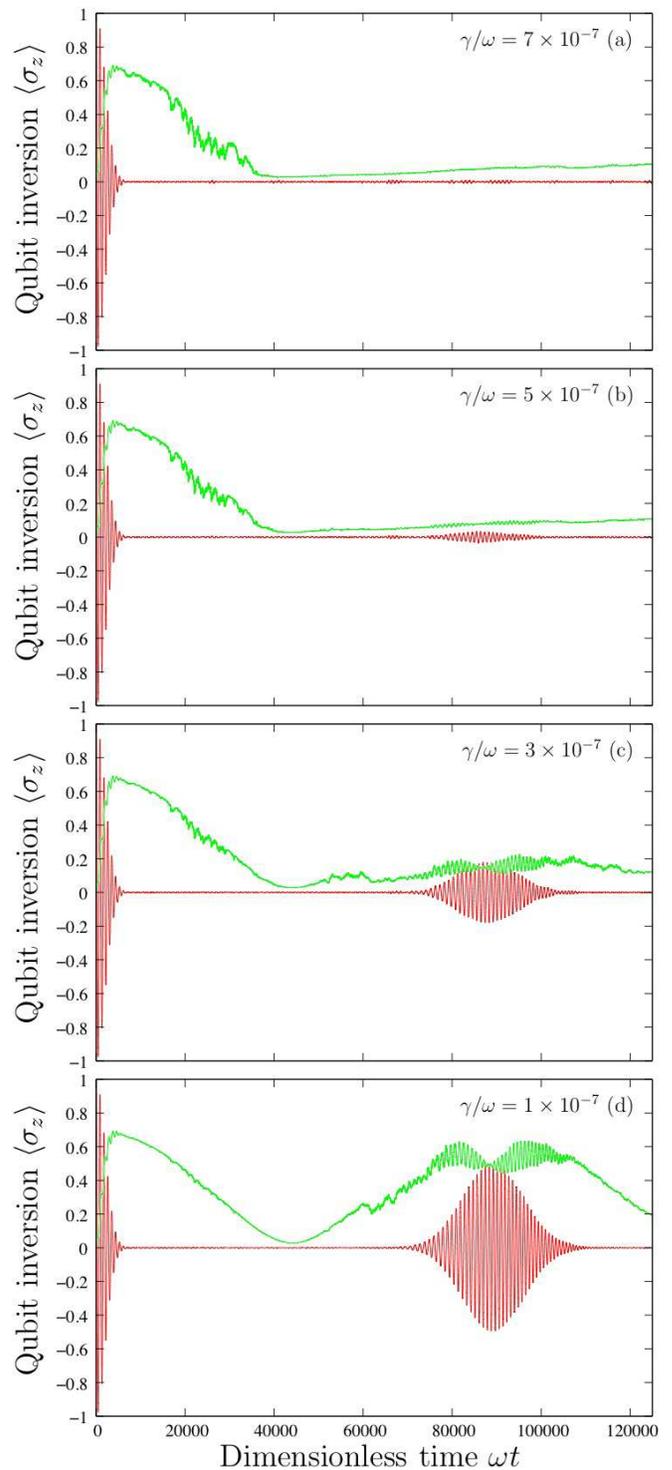}}
\end{center}
\caption{(Color online) Results (dark grey/red) showing 
qubit inversion, $\langle \sigma_z \rangle$,
for different values of dissipation (and therefore drive) applied to the field.
These plots focus on the parameter range where the revival of Rabi 
oscillations begins to emerge. For each
individual QSD run shown, the qubit entropy in Nats (light grey/green) is superimposed.} \label{fig3}
\end{figure}

In the quantum limit, the bottom right of figure~\ref{fig2}, the quantum noise in the
single run of QSD is so small that the entropy of the qubit in this single run gives
a very reasonable approximation to the entropy of the ensemble, which matches well with the
qubit entropy in the exactly zero dissipation limit. Figure~\ref{fig3} gives focus to 
the parameter range where the revival begins to emerge.
For all the individual QSD runs that are neither well into the classical or well into the
quantum limit, we still plot the qubit entropy for that run. It should be understood
that these do not approximate well the ensemble entropy in this crossover parameter
regime. Nevertheless, as the damping and drive applied to the field are reduced,
these plots do illustrate the emergence of an entropy peak sympathetic with the collapse,
an entropy dip at the ``attractor time''~\cite{geaban90,geaban91} (half the revival time),
and the emergence of entropy oscillations sympathetic with the revival.

In conclusion, we have shown how to take a quantum field through the
quantum-classical crossover, as seen through the behaviour of a
coupled qubit, used as a probe. Crucially, this crossover is not
made through increasing the field strength; rather, it is made at a
fixed field strength, through adding dissipation to the field. So
the qubit Rabi frequency is fixed through the whole range of
behaviour shown. Thus a quantum field with around fifty photons
in it can be made to act rather classically, or still show its
characteristic quantum properties, dependent upon the environment
coupled to it. Perhaps amusingly, the apparent coherence of the
probe qubit is {\it increased} by {\it adding} dissipation to the
field to which it is coupled. This is demonstrated through increased
persistence of the qubit Rabi oscillations, or suppression of the
collapse. Our work provides an explicit example of how it is
possible to decohere and make classical part of a coupled system, but---crucially---without 
that decoherence leaking through to decohere the other,
still quantum, part of the system. The classical part of the system loses its 
ability to entangle with the still quantum part, but in this case does not
destroy the coherence of the probe qubit. As stressed in the introduction,
the model used here is relevant for a broad range of
physical systems. To fully explore the
quantum-classical crossover in experiments, the ``intrinsic''
coherence time of the probe qubit (limited by all other environment
effects, not its coupling to the field) must exceed the
revival time. This is likely to be very challenging in current
experiments. However, evidence for the ``suppression of collapse''
only requires the intrinsic qubit coherence time to exceed the
collapse time. This is much less demanding, and provides an initial
goal for current qubit-field experiments of various kinds.

\begin{acknowledgments}We thank Catherine Jarvis for helpful discussions.\end{acknowledgments}

\end{document}